\documentclass[12pt]{article}
\usepackage{pdproc} 

\usepackage{epsfig}
\usepackage{amsmath}
\usepackage{amssymb}
\usepackage{amsfonts}
\usepackage{mathrsfs}
\usepackage{color}

\newcommand{\la}{\lambda}

 \def\cO{{\cal O}}

\newcommand{\beq}{\begin{equation}}
\newcommand{\eeq}{\end{equation}}
\newcommand{\bac}{\beq\begin{array}}
\newcommand{\eac}{\end{array}\eeq}
\newcommand{\ba}{\begin{array}}
\newcommand{\ea}{\end{array}}
\newcommand{\bea}{\begin{eqnarray}}
\newcommand{\eea}{\end{eqnarray}}

\newcommand{\sign}{\mathrm{sign}}

\begin{document}

\renewcommand{\thefootnote}{\alph{footnote}}

\mathversion{bold}
\title{ A $T'$ FLAVOUR MODEL FOR FERMIONS AND ITS PHENOMENOLOGY}
\mathversion{normal}

\author{LUCA MERLO}

\address{Physik-Department, Technische Universit\"at M\"unchen\\
Institute for Advanced Study, Technische Universit\"at M\"unchen\\ 
Lichtenbergstr. 2a, D-85748 Garching, Germany\\
 {\rm E-mail: luca.merlo@ph.tum.de}}

\abstract{We present a supersymmetric flavour model based on the $T'$ discrete group, which explains fermion masses and mixings. The flavour symmetry, acting in the supersymmetric sector, provides well defined sfermion mass matrices and the resulting supersymmetric spectrum accounts for sufficiently light particles that could be seen at LHC. Furthermore, several contributions to FCNC processes are present and they can be useful to test the model in the present and future experiments. We will review the main results for both leptons and quarks.}
   
\normalsize\baselineskip=15pt

\section{Introduction}
The flavour sector of the Standard Model is problematic. Indeed there is no explanation for the strong mass hierarchies in the charged fermion sector, for the milder mass hierarchies in the neutrino sector, for the small mixing angles of the CKM matrix and for the (two) large and one almost vanishing mixing angles in the PMNS matrix. In particular the latter is likely described by simple numbers in the so-called Tri-Bimaximal (TB) pattern\cite{TB}, which corresponds to the mixing angles $\sin^2\theta^{TB}_{12}=1/3$, $\sin^2\theta^{TB}_{23}=1/2$, $\sin\theta^{TB}_{13}=0$, and agrees at the $1\sigma$ level with the data.

Flavour symmetries can help with this regards. A lot of effort has been devoted to construct models in which the Lagrangian of the theory is invariant under a specific flavour symmetry at a flavour scale equal or higher than the electroweak one. Since at lower energy scales no explicit symmetry can be deduced by the Yukawa matrices, a flavour symmetry breaking mechanism should be implemented in such models. 

Usually, the introduction of a flavour symmetry and of its breaking mechanism corresponds to the introduction of New Physic (NP) particles, which could however mediated directly or indirectly new contributions to FCNC observables.

It has been shown\cite{MFV} that the Minimal Flavour Violation (MFV) approach, based on the flavour symmetry $U(3)^5$, is powerful to suppress any potentially dangerous FCNC contribution arising from NP. However, a natural explanation for the fermion masses and mixings in this context is still missing \cite{AGMR:MFVPotential}. Furthermore, the spontaneous breaking of a global continuous symmetry would produce the appearance of unobserved Goldstone bosons. An elegant solution to this problem is to gauge the symmetry\cite{MFVgauged}, but dangerous NP contributions arise from the modification of the SM couplings and from new diagrams mediated by the new neutral gauge bosons\cite{BMS:BSG}.

The difficulties in the MFV context suggest that the $U(5)^3$ flavour symmetry is probably too large and restrictive to allow for a natural solution to the flavour problem. However, there is a large variety of possible choices for such flavour symmetries: either global or local, either Abelian or non-Abelian, either continuous or discrete.

In a series of papers\cite{A4} it has been pointed out that a broken flavour symmetry based on the discrete group $A_4$, the group of even permutations of four elements, appears to be particularly suitable to reproduce the TB lepton mixing pattern as a first approximation. In the following we present a model based on the discrete group $T'$\cite{Tp}, which extends the $A_4$ description of leptons to the quark sector, getting a realistic CKM matrix. We then discuss the corresponding phenomenological analysis.

\mathversion{bold}
\section{The $T'$ flavour model}
\mathversion{normal}

We recall here the main features of the $T'$ model\cite{Tp}. The flavour group $G_f$ is given by the product of $T'$ and other factors: the spontaneous breaking of $T'$ is responsible for the TB mixing, while the other factors help neglecting unwanted couplings and providing the correct hierarchy among the charged fermions. 

The TB mixing is achieved through a well-defined symmetry breaking mechanism: $T'$ is not broken completely, but two groups remain unbroken, i.e. a $Z_2\times Z_2$ in the neutrino sector and a $Z_3$ in the charged lepton sector. This mechanism is due to a set of scalar fields, the flavons, which transform only under $G_f$ and break the flavour symmetry developing specific vacuum expectation values (VEVs). 

Once defined the transformation properties of all the fields, we can write an effective Lagrangian of non-renormalizable operators in terms of matter fields and flavons, suppressed by powers of the flavour scale $\Lambda_f\approx 10^{16}$ GeV. Once the flavons develop VEVs, masses and mixings arise in term of a new parameter $u=VEV/\Lambda_f$ which spans the range $[0.007,\,0.05]$. 

At the leading order (LO), the charged leptons get a diagonal mass matrix with hierarchical entries, while the neutrino mass matrix, arising through the Weinberg operator, is diagonalized by the TB matrix. The LO structure of the lepton mixing matrix is then the TB pattern. When the higher order operators of the effective theory are considered, corrections affect the mass matrices and translate into deviations from the TB values of the lepton mixing angles: $\sin^2\theta^\ell_{12}\simeq1/3+\cO(u)$, $\sin^2\theta^\ell_{23}\simeq1/2+\cO(u)$, $\sin\theta^\ell_{13}\simeq\cO(u)$.

Considering the quark sector, the particular choice of the transformation properties for the quark fields and the specific alignment of the flavon VEVs determine a ``shell'' filling of the mass matrices. Reporting the NLO results in terms of the Cabibbo angle $\lambda\approx0.23$, we get
\beq
M_u\propto\left(
        \begin{array}{ccc}
           \la^8  & \la^6 & \la^6 \\
            \la^6 & \la^4 & \la^4 \\
            \la^4 & \la^2 & 1 \\
        \end{array}
\right)\qquad
M_d\propto\left(
        \begin{array}{ccc}
           \la^8  & \la^5 & \la^6 \\
            \la^5 & \la^4 & \la^4 \\
            \la^6 & \la^4 & \la^2 \\
        \end{array}
\right)\;.
\eeq
To find this result, which gives realistic mass hierarchies and mixings in the quark sector, we invoked a small fine-tuning of order $\lambda$. It is interesting to note that, despite of the large number of unknown parameters, the model present two distinct predictions in the quark sector: 
\beq
\sqrt{\dfrac{m_d}{m_s}}=\left|V_{us}\right|+\cO(\lambda^2)\,,\qquad\qquad \sqrt{\dfrac{m_d}{m_s}}= \left|\dfrac{V_{td}}{V_{ts}}\right|+\cO(\lambda^2)\,.
\label{p1}
\eeq
The first one is the well-known Gatto-Sartori-Tonin relation between masses and mixing angles, which results in good agreement with the data; the second one needs some additional corrections.

\section{The phenomenological analysis}

\begin{figure}
\centering
\leftline{\hfill\vbox{\hrule width 5cm height0.001pt}\hfill}
\includegraphics[height =12cm]{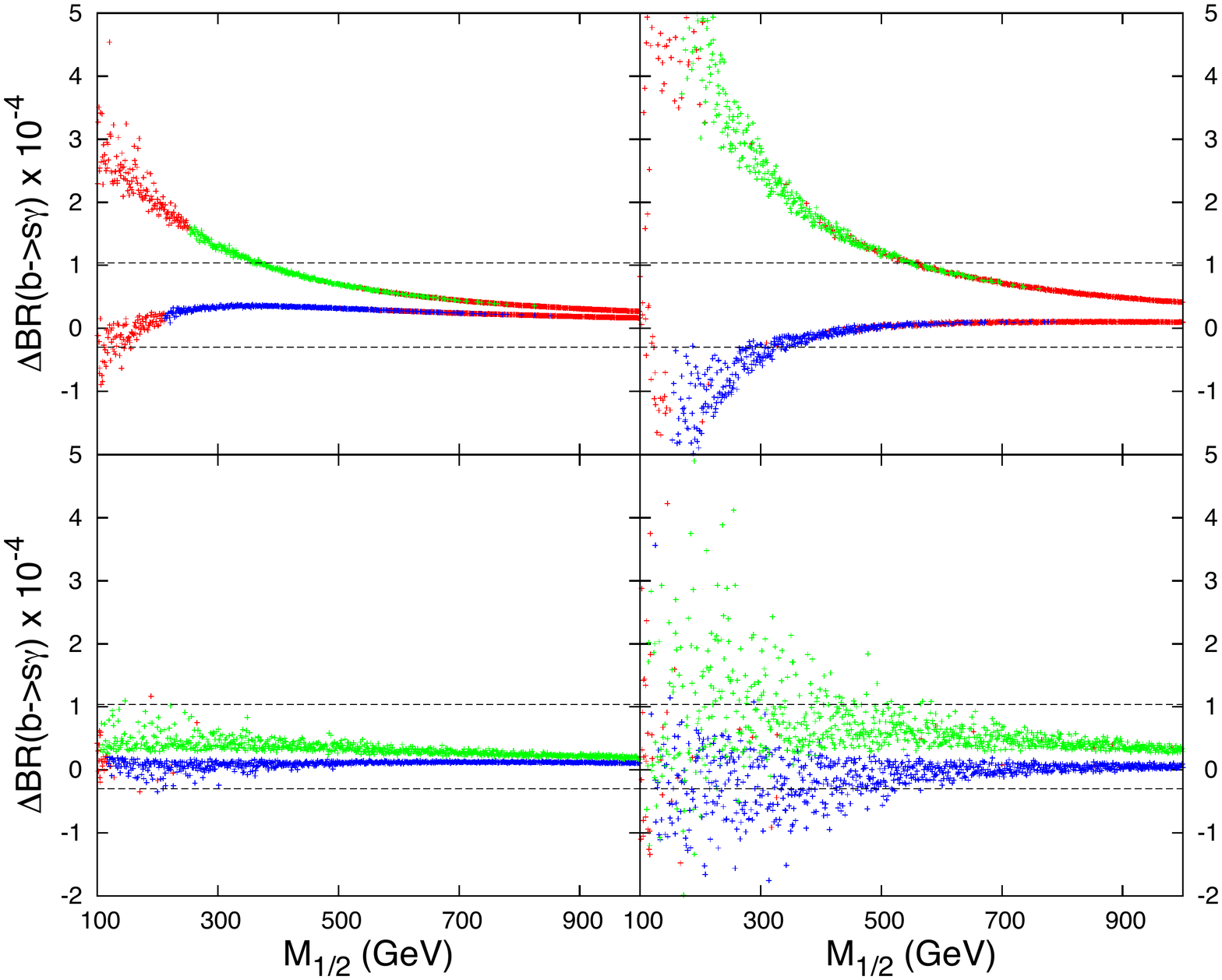}
\leftline{\hfill\vbox{\hrule width 5cm height0.001pt}\hfill}
\caption{$BR(b\to s\gamma)$ as a function of $M_{1/2}$. Blue (Green) points stand for $\mu>0$ ($\mu<0$), while the red points represents excluded values of the parameter space, due to the lightest chargino mass bound. The dashed lines correspond to the values reported in the text.}
\label{picture1} 
\end{figure}

\begin{figure}
\centering
\leftline{\hfill\vbox{\hrule width 5cm height0.001pt}\hfill}
\includegraphics[height =12cm]{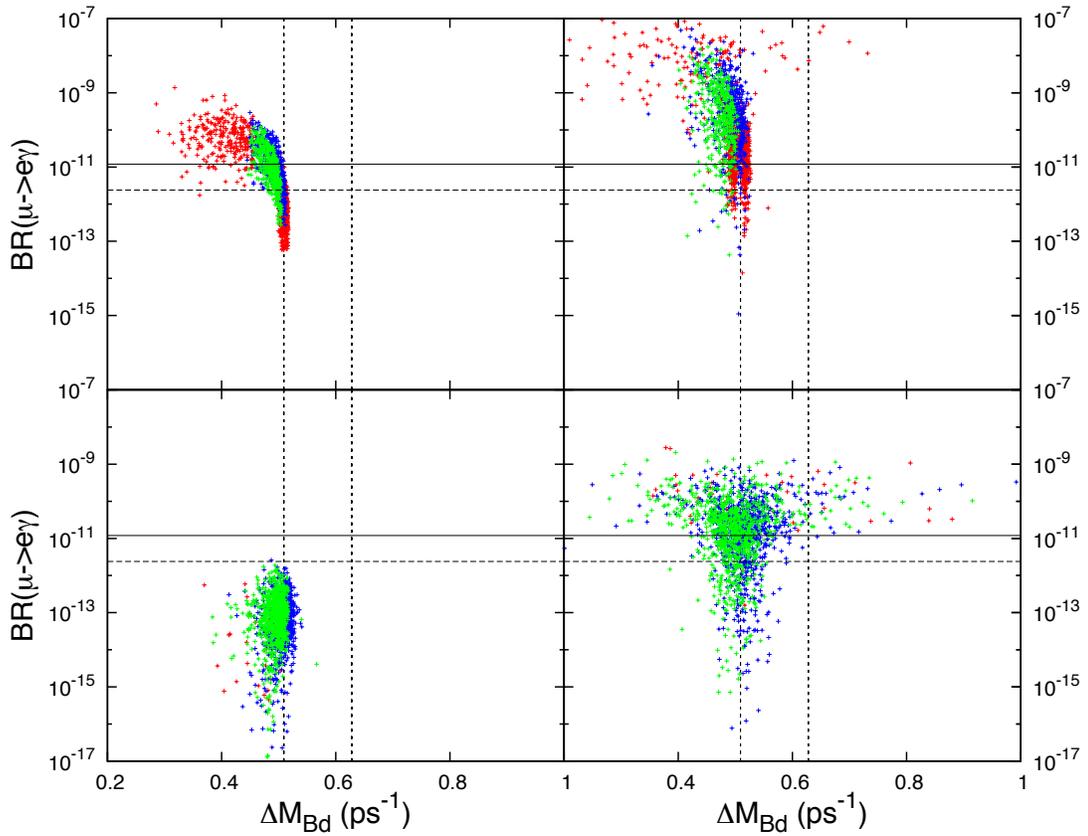}
\leftline{\hfill\vbox{\hrule width 5cm height0.001pt}\hfill}
\caption{Correlation among $BR(\mu\to e\gamma)$ and $\Delta M_{B_d}$. Blue (Green) points stand for $\mu<0$ ($\mu>0$), while the red points represents excluded values of the parameter space, due to the lightest chargino mass bound. The dashed lines correspond to the values reported in the text.}
\label{picture2}
\end{figure}

The analysis on the lepton sector is similar to that one of the lepton $A_4$ model\cite{FHLM_LFV}. The few differences in the VEV alignment and in the particle spectrum have negligible effects on the results. We then only focus on the quark sector, showing few results of the analysis performed in\cite{MRZ:Tp}. In Fig.~(\ref{picture1}), we show scatter plots for $BR(b\to s\gamma)$ as function of the supersymmetric gaugino mass $M_{1/2}$.  In Fig.~(\ref{picture2}), we show the correlation among the $BR(\mu\to e\gamma)$ and the mass splitting in the $B_d$ system, $\Delta M_{B_d}$, which turn out to be the most sensitive observables.

In all the scatter plots, the free parameters of the models are random real numbers in the range $[1/2,\,2]$; $\tan\beta=5(15)$ and $u=0.01(0.05)$ in the plots in the left (right); the common scalar mass $m_0$ and the soft trilinear terms $A_0$ are fixed to $A_0=2\,m_0=400(2000)$ GeV in the upper (lower) plots. Red points are excluded by imposing ``phenomenological'' requirements on the Higgs and/or SUSY masses. Blue (Green) points refer to $\mu > 0$ ($\mu < 0$). 

In Fig.~(\ref{picture1}) we show the ``differential'' Branching Ratio for the $\bar B \to X_s \gamma$ decay,
\beq
\Delta BR(\bar B \to X_s \gamma) = BR_{SM+NP}(\bar B \to X_s \gamma) - BR_{SM}(\bar B \to X_s \gamma)\,,
\eeq
and we compare it with the current experimental value \cite{PDG2010}, measured with a photon--energy cut-off 
$E_\gamma > 1.6$ GeV in the $B$-meson rest frame,
\beq
BR(\bar B \to X_s \gamma) = (3.55 \pm 0.24 \pm 0.09) \times 10^{-4} \,,
\eeq
subtracted by the SM prediction calculated at NNLO \cite{BSGNNLO} for the same 
photon energy cut-off,
\beq
BR(\bar B \to X_s \gamma) = (3.15 \pm 0.23) \times 10^{-4} \,.
\eeq
In the plots, the horizontal dashed lines represent the $2\sigma$ experimental bound from Ref.~\cite{PDG2010}. 
In the small $(A_0,m_0)$ region (upper plots), the dominant $T'$ contribution to the $b \to s \gamma$ BR is 
typically the charged Higgs one, due to the fact that the stop is for most of the $M_{1/2}$ range heavier than the 
$H^\pm$ (and the chargino heavier than the top). The Higgs contribution is always concordant in sign with the SM one. 
The second most relevant contribution is the chargino one with a sign depending on $\sign[\mu]$: it tends to enhance 
(cancel) the SM contribution for $\mu < 0$ ($\mu > 0$). Gluino contributions are practically independent 
from $\sign[\mu]$, while neutralino ones are completely negligible. As a consequence, experimental constraints on the 
$b \to s \gamma$ BR tend to disfavor $\mu < 0$ for $M_{1/2}\lesssim500$ GeV especially for the larger $\tan\beta$ values (right plots). The positive $\mu$ case, instead, turns out to be mostly allowed. In the $A_0=2\,m_0=2000$ 
GeV scenario (lower plots) all new physics contributions get strongly suppressed and no significative limits are 
expected from the $b \to s \gamma$ measurement in neither the two $\tan\beta$ cases considered.

In Fig.~\ref{picture2} the $T'$ correlation between $BR(\mu \to e \gamma)$ and
$\Delta M_{B_d}$ is shown varying the common gaugino mass $M_{1/2}$ in the range 
$(100,1000)$ GeV. The horizontal (solid) dashed lines indicate the $2\sigma$ experimental bounds as reported below:
\beq
BR(\mu\to e \gamma) = (1.2\times 10^{-11})\, 2.4 \times 10^{-12}\,,\qquad\qquad
\Delta M_{B_d} = 0.507 \pm 0.005 \quad {\rm ps}^{-1}\,.
\eeq
It is evident from Fig.~\ref{picture2} that the strongest constraint comes from the $BR(\mu \to e \gamma)$, which almost excludes 
the small $m_0$ region. However, when larger values of $m_0$ are considered (lower plots), an improvement in the 
$BR(\mu \to e \gamma)$ bound will not impose any severe exclusion on the model, as BR values as small as $10^{-15}$ are still acceptable. Some help can come, instead, from the hadronic sector. An improvement in the knowledge of the meson mass differences at future facilities can help in further constraining the model in the large $\tan\beta$ regime (see lower--right plot of Fig.~\ref{picture2}), while the large $m_0$, small $\tan\beta$ case will be almost impossible to exclude.

A complete analysis\cite{MRZ:Tp} of these and other observables in the full parameter space reveals that relatively light sparticles are allowed, which could be found even at LHC. This would provide a good test of the model.

\section{Acknowledgements}
It is a pleasure to warmly thanks the organizers of ``WONP-NURT 2011'' (February 7--11, 2011, Havana, Cuba), of ``The XIV International Workshop on Neutrino Telescopes'' (March 15--18, 2011, Venice, Italy), and of ``PLANCK 2011 - From the Planck Scale to the ElectroWeak Scale'' (May 30 -- June 3, 2011, Lisbon, Portugal) for giving us the opportunity to report on our recent work. Furthermore I would like to thank Stefano Rigolin and Bryan Zald\'ivar Montero for the fruitful collaboration.


\end{document}